\documentclass[12 pt]{article}
\usepackage{amsmath, amsfonts, amssymb}
\usepackage{graphicx}
\usepackage{indentfirst}
\usepackage{threeparttable}
\usepackage{url}
\usepackage{soul}
\usepackage{natbib}
\usepackage{aas_macros}

\newcommand{\hb}{\\ \hspace*{2ex}}

\begin{document}

\title{\LARGE \bf Interpretation of ALMA velocity map for the obscuring torus in  NGC1068}
\author{\bf E.~Yu.~Bannikova $^{1,2}$\thanks{E-mail: bannikova@astron.kharkov.ua},  
N.~A.~Akerman$^{2}$, A.~V.~Sergeyev$^{1,2}$}
\date{\it  \small  $^1$ Institute of Radio Astronomy of National Academy of Science of Ukraine, \hb 
Mystetstv 4, UA-61022 Kharkiv, Ukraine, \\
$^{2}$ V.N.Karazin Kharkiv National University, \hb Svobody Sq.4, UA-61022,  Kharkiv, Ukraine}

\maketitle

\renewcommand{\abstractname}{}
\begin {abstract}
\bf Abstract \rm  

Recent ALMA observations have resolved the obscuring torus in the nearest Sy2 galaxy, NGC1068, in the millimeter band. These observations have confirmed the presence of a geometrically thick torus with an orbital motion of its matter and the velocity distribution which can reflect the clumpy structure. In the framework of N-body simulations we consider a dynamical model of an obscuring torus which accounts for the gravitational interaction between the clouds moving in the field of the central mass. In considered model, clouds are orbiting around the central mass exhibiting a spread in inclination and eccentricity. The self-gravity of the torus induces the velocity distribution of clouds with a global orbital motion which mimics the ALMA data for NGC1068.

\end{abstract}


\section{Introduction}

The unified scheme of AGNs is the paradigm accounting for the differences in AGN types (Antonucci 1984, Antonucci \&  Miller 1985, Urry \& Padovani 1995). The key feature of this scheme is a dusty torus located around an accretion disk in the central region of the AGN. Type II AGN occurs when the torus is located edge-on and the emission of the accretion disk together with the Broad Line Region (BLR) is obscured by the dust in the torus. On the contrary, Type I shows up when the torus is inclined at some angle with respect to an observer. In order to satisfy all the statistical data for numerous Sy galaxies, a conclusion was made about its geomatrically thick structure (Osterbrock \& Shaw 1988).

One important question was whether the matter distribution in the torus was continuous or clumpy. The first suggestion of a clumpy structure of the dusty torus was made by Krolik \& Begelman (1988). Indeed, the temperature for the continuous matter with a high orbital motion would be too hot for the dust to survive. The IR Spectral Energy Distributions (SEDs) of Sy galaxies satisfy the assumption that the torus has a clumpy structure with a Gaussian distribution of clouds in its cross-section (Nenkova et al. 2008, Schartmann et al. 2010, Garcia-Gonzalez et al. 2017). So, the dusty torus of an AGN looks like a toroidal obscuring region (Nenkova et al. 2008) with a smooth distribution of clouds. Such a structure can provide us some information about the conditions of torus formation. 

The rotation curves of the matter in the torus of the nearest Sy2 galaxy NGC1068 were obtained by mapping the H$_2$O megamaser spots with different interferometers:  VLBI (Greenhill et al. 1996), VLA (Gallimore et al. 1996), and VLBA (Gallimore et al. 2004). The results of these observations showed that the megamasers spots are distributed along a chain in the range of the scales 0.4-0.6 pc with corresponding velocities 250-300 km/s.  The high velocities of the clumps and the conditions of pumping of the H$_2$O megamasers prove that this emission forms on the inner edge of the torus. It was also found out that the rotation curve is not Keplerian, which demonstrates the possible influence of the torus self-gravity. The interpretation of the rotation curves from the observation of the water megamaser radiation in the torus of NGC1068 in the framework of the Keplerian disk motion gives an estimation of the supermassive black hole (SMBH) mass which is comparable to the disk mass  (Lodato \& Bertin 2003, Hur\'e 2002). However, our N-body simulations  show that the very massive torus with $M_{torus} \approx M_{BH}$ will be unstable and flow out during a few orbital periods (the paper in preparation).  

The first direct observations of the torus in NGC1068 were carried out with the help of VLTI/MIDI in the IR band (Jaffe et al. 2004), with the following main results. The torus is geometrically thick and has parsec  scales. There are two regions in the temperature distribution. The hot region \mbox{(T$>800$K)} seems to be the result of the clouds being heated by the radiation of the accretion disk; the warm region \mbox{(T$=320$K)} is the main torus body with the outer edge of about 3pc. The temperature stratification supports the idea of a clumpy structure in the torus. Similar results were also obtained for another Sy2 galaxy Circinus A (Tristram et al. 2007, Tristram et al. 2014).

The next step in the investigation of the obscuring tori was made with the help of ALMA in the millimeter band (Garcia-Burillo et al. 2016, Imanishi et al. 2018). The first result for NGC1068 was a CO molecule emission which came out of the torus with the diameter about 10pc and with the major-to-minor radius ratio equal to 0.5. The orientation of the torus coincides with that of the S1 radio source which was found in (Gallimore et al. 1996, 2004). The mass of the gas in the torus was also estimated to be $M_{gas} \approx 10^5 M_\odot$. The next cycle of the ALMA observation was made with better resolution, in the molecules HCN and HCO$^+$ (Imanishi et al. 2018), and one of the unexpected results was the estimation of the SMBH mass. The observational velocity is 20km/s at the distance 3pc (Imanishi et al. 2018) and, having taken into account the torus inclination angle $i=34^\circ$, the orbital velocity was found out to be 36km/s. In the framework of the Keplerian motion, it means that the mass of the SMBH is $9 \times 10^5 M_\odot$, which is much less that the mass derived from the previous results (VLA, VLBI). So, there is a disagreement between the estimation of SMBH mass derived from the megamaser radiation and the ALMA observations, which may suggest that the approach of a Keplerian disk does not work. Indeed, the observations demonstrate that the torus self-gravity is needed to be taken into account as well as a more complicated motion of the matter in the torus, a thick toroidal shape, a clumpy structure, etc. It means that we need to consider the dynamical model of the torus that is located in the gravitational field of the central mass. Such a model was suggested in (Bannikova et al. 2012) and further developed in (Bannikova 2015, Bannikova \& Sergeev 2017). Here we present the work in progress using the dynamical model of the torus for an  explanation of the ALMA observations of the velocity distribution.

\section{Dynamical model of a dusty torus}

We consider a dynamical model of the dusty torus in AGNs in which the toroidal structure is formed by the clouds moving in the gravitational field of the central mass with $M_{BH}$ (Bannikova et al. 2012).  In such a model we include the self-gravity of the torus with mass $M_{torus}$ in the framework of the N-body problem. It means that we are taking into account the gravitational interaction between each cloud of the system. In our previous work it was shown that the gravitating torus in the field of the central mass is stable and keeps its geometrically thick shape for the different initial conditions (Bannikova 2015, Bannikova \& Sergeyev 2017). It is only important that in the initial state the distribution of clouds has a spread in inclinations and in eccentricities. The numerical experiments demonstrated that the torus is stable if its mass is less than 10\% of the central mass.
To explain the ALMA observation of the velocity distribution, we choose the initial conditions for the N-body simulation using the idea of the torus formation. In the pre-active stage of an AGN the clouds can move around the central mass in orbits with random distribution of all the elements. The toroidal structure can be formed as a result of the beginning of the active stage of an AGN, when the radiation pressure from the accretion disk ejects clouds, forming two wind cones.

In order to implement this idea, we use the corresponding initial conditions with an half-opening angle of the wind $45^\circ$. We have chosen this value of the opening angle according to the spectroscopy data of ionization cones for NGC1068 (Das et al. 2006) and the recent VLTI/GRAVITY observations of BLR for 3C273 (Gravity Collaboration 2018). It means that inclinations ($i$) of all clouds are in the range [$0,\pi/4$]. The others orbital elements: semi-major axis ($a$), eccentricity ($e$), longitude of the ascending node ($\Omega$), argument of periapsis ($\omega$) are distributed randomly. We use the central mass-to-torus mass ratio as a parameter in N-body simulation and demonstrate the result for $M_{torus}=0.01 M_{BH}$. We choose this value to coincide it with the estimation of gas in the torus from ALMA observations: $M_{gas} \approx 10^5 M_\odot$ (Garcia-Burillo et al. 2016, Imanishi et al. 2018) and with the suggestion that SMBH mass is $M_{BH} \approx 10^7 M_\odot$.  Note, that the SMBH mass is complicated to obtain directly from the observations of Sy2 galaxies because in this case the central region of AGN is hidden by the dusty torus. Thus, the SMBH mass is model dependent. As it was mentioned above, the analysis of the megamaser radiation in NGC1068 gave $M_{BH} \approx 10^7 M_\odot$ (Lodato \&Bertin 2013, Hur\'e 2002), but this value was obtained without an accounting of the features of clouds motion in the torus due to self-gravity. It means that in general the SMBH mass can be a parameter in our model and can be derived from the comparison of the results of a simulation and the observed velocity distributions (see next Section).
\begin{figure}[!h]
\centering
    \includegraphics[width=0.4\textwidth]{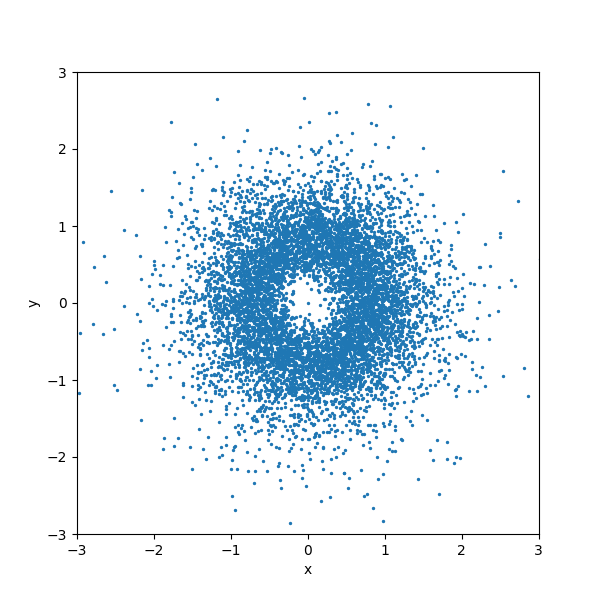}
    \includegraphics[width=0.4\textwidth]{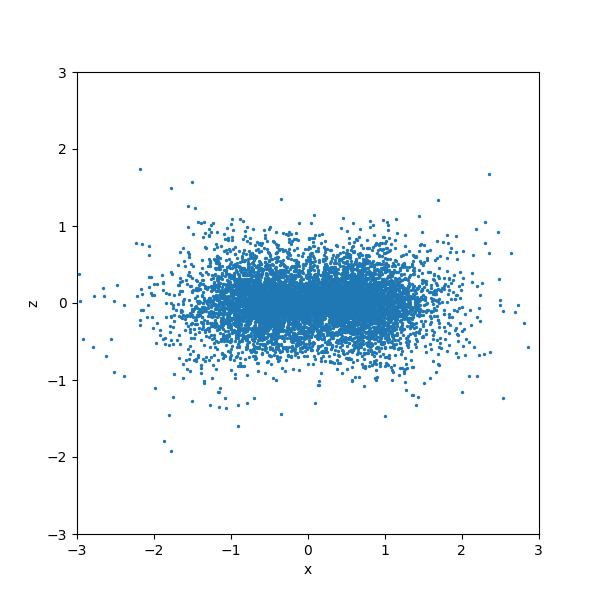}
    \includegraphics[width=0.4\textwidth]{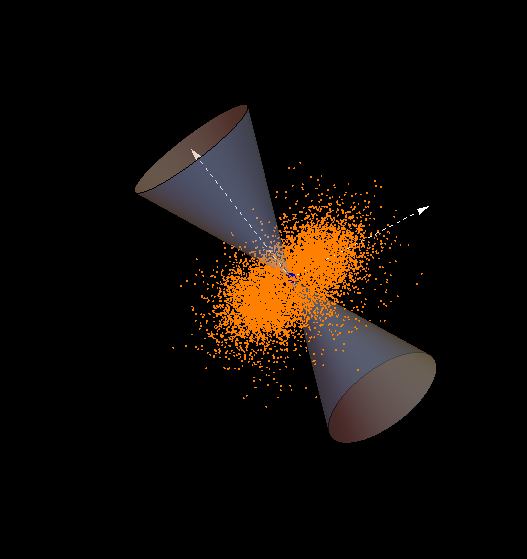}
      \caption{The resulting distribution of N-body simulation of the torus in the gravitational field of the central mass. {\it Left}:  projection on the equatorial plane; {\it right}: projection on the meridional plane; {\it bottom}: 3D distribution of the clouds with wind cones.  Initial conditions: $M_{torus}=0.01M_{BH}$, $N=8192$.}
    \label{fig:distr}
\end{figure}
The other parameter of the model is the number of clouds which can be chosen to satisfy the condition of obscuration: $N=10^5$ (Nenkova et al. 2008). From the other point, it was shown in (Bannikova et al. 2012) that we can predict the behavior of the system for different number of particles. Indeed,  the total change of a particle velocity due to random nature of its gravitational interactions is $(\triangle V)^2 \sim M_{torus}^2 \triangle t/ N$. It means that the same $(\triangle V)^2$ and the same torus shape in equilibrium state will be reached for a larger number $N_2$ during longer time interval of the simulation $\triangle t_2 =\triangle t_1 N_2/N_1$.   Here we use $N=8192$ to demonstrate the distribution of the clouds and to analyze their dynamics.  

\begin{figure}[!h]
    \centering
    \includegraphics[width=0.3\textwidth]{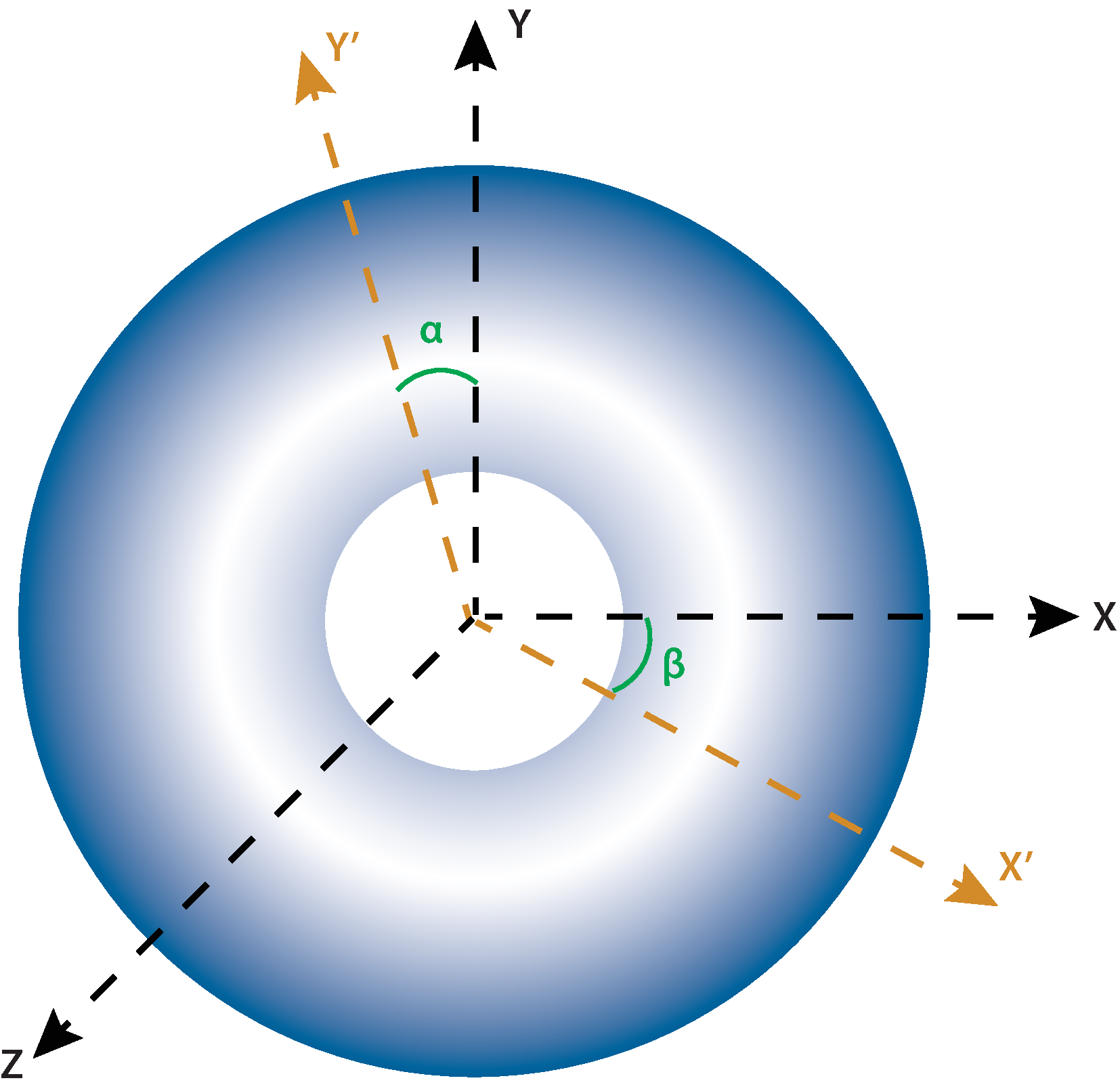}
    \caption{The scheme of a torus showing two rotation angles: $\alpha$ is around the $x$-axis and $\beta$ is around the $y$-axis}
    \label{fig:rotation}
\end{figure}

The results of the simulation demonstrate that the torus changes its cross-section due to self-gravity during the first orbital periods and then reaches its equilibrium state.  The resulting distribution of clouds in the torus is presented on Fig.\ref{fig:distr}. This distribution obeys the Gaussian law and satisfies the obscuration condition which is needed to explain the observational IR SED (Nenkova 2008). The clouds are distributed in the toroidal region with the spread in velocities which can be reflected in the velocity map in mm band taking into account the torus orientation (see Fig.\ref{fig:rotation} and an explanation below).

\section{Dynamics of the clouds in the torus and interpretation of ALMA results}

\begin{figure}[!h]
    \centering
    \includegraphics[width=0.35\textwidth]{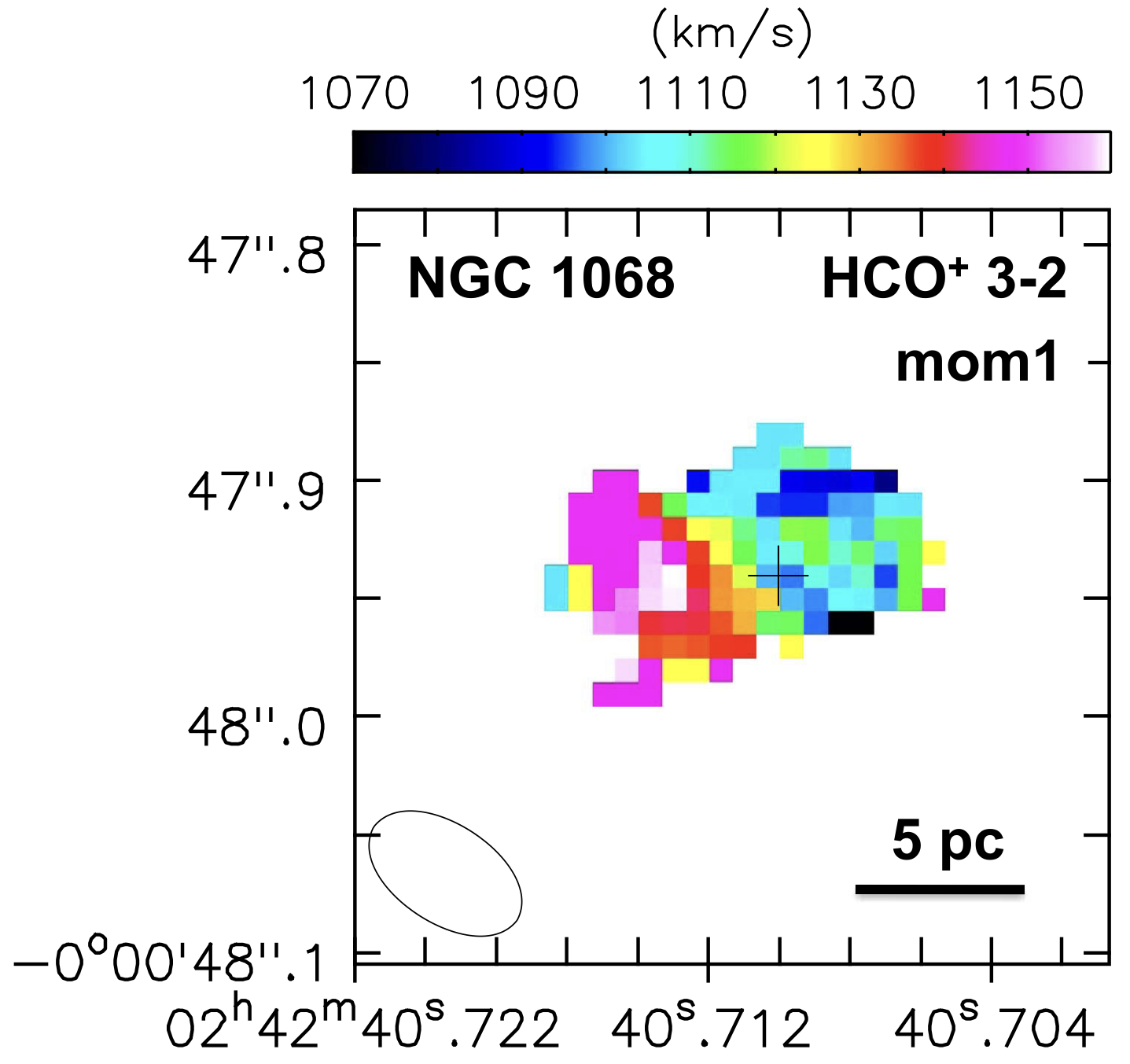}
    \includegraphics[width=0.3\textwidth]{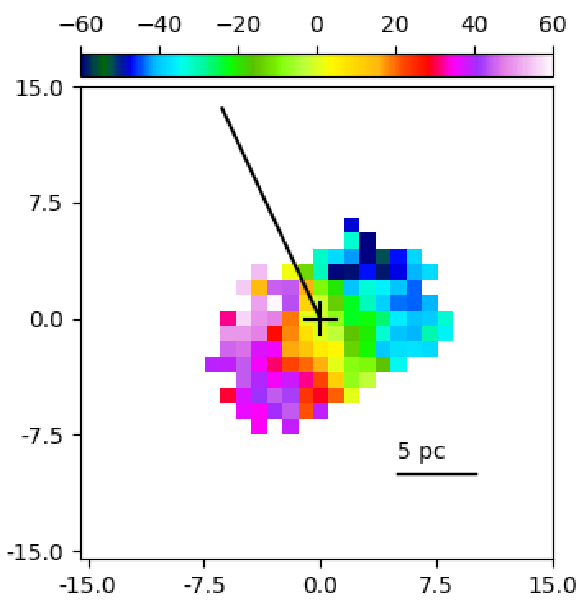}
    \includegraphics[width=0.3\textwidth]{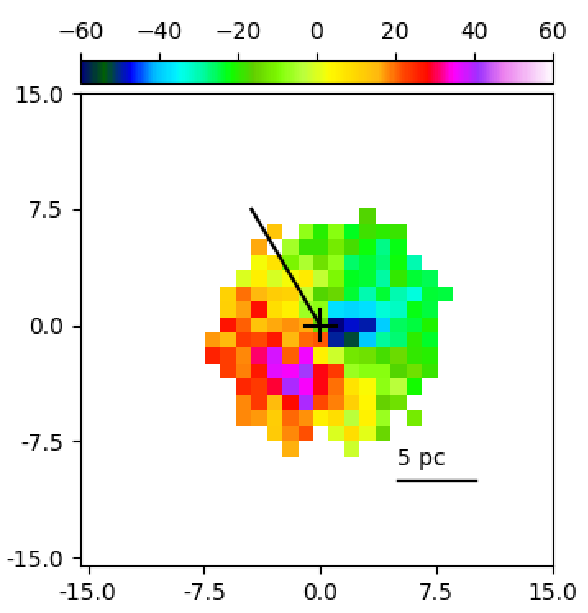}
    \caption{The distribution of mean velocity in the torus of NGC1068. {\it Left}: 
        ALMA observations of HCO$^+$ -- picture from (Imanishi 2018). {\it Middle}: the result of N-body simulation of the torus which is turned by two angles  $\alpha=65^\circ$ and $\beta=90^\circ$. {\it Right}: the same simulation but with the angles  $\alpha=30^\circ$, $\beta=20^\circ$. Black crosses indicate the central mass position and black lines indicate the torus symmetry axis. Initial condition corresponds to Fig.\ref{fig:distr}.}
    \label{fig:vel_map}
\end{figure}

As it was mentioned above, the ALMA observations of the central region of NGC1068 allow to understand the dynamics of the matter in the obscuring torus. The mean velocity distribution of the ALMA data (Fig.\ref{fig:vel_map}; left) clearly shows the global rotation of the matter (colored with red and blue) as well as some inhomogeneous spots indicating that the torus has a clumpy structure. To interpret these observations, we have to take into account the orientation of the torus in such a way  that the symmetry axes coincides with the cones direction (Das et al. 2006).  It is also necessary to take into account the orientation of the torus with respect to the line of sight. Since the distribution of clouds in the torus cross-section is a Gaussian one, then for a certain orientation of the torus (almost from the edge-on), the number of clouds along the line of sight is large and the external clouds can obscure the central ones. For a different spatial orientation of the torus (near to face-on), faster clouds that are near the accretion disk can be accessible to the observer and will be displayed on the velocity map  as regions with a higher values.
 
To determine the orientation of the torus, we choose the reference system in such a way that the $z$-axis coincides with the line of sight and with the torus symmetry axis (Fig.2). So, the resulting velocity map is the projection on $xy$-plane ($x$ is horizontal axis).  Since the torus is an axisymmetrical body, only two Euler angles are required. The angle $\alpha$ determines the orientation of the torus by edge-on/face-on ($\alpha =90^\circ$ corresponds to edge-on). The angle $\beta$ determines the turn around $y$-axis. The observing angle is the result of these two rotations.  To construct the velocity maps from our N-body simulation (Fig.\ref{fig:distr}) we use  the following algorithm.

1. Turn the torus by two angles such a way that the axis symmetry will coincide with observed wind cones orientation.

2. Take into account that the clouds closer to the observer will be seen first.
     
3. Divide the plot by $30\times30$ cells so that the cell scale matches the ALMA velocity map.
          
4. Find the mean velocity of the clouds which are located in the cylinder along the line of sight with the cross-section corresponding to the cell but taking into account only the first 10 clouds as they fully obscure the others.

The model velocity maps for different orientations of the torus which correspond to the same projection of the symmetry axis are presented on Fig.\ref{fig:vel_map} ({\it middle}, {\it right}). It can be seen that the resulting velocity maps depend on the orientation of the torus.  Fig.\ref{fig:vel_map} {\it middle} demonstrates good agreement with the observations (Fig.\ref{fig:vel_map} {\it left}). So, the velocity spread in the cells reflects the spread in clouds velocities in the torus due to its self-gravity.  Fig.\ref{fig:vel_map} {\it right} has some differences in the shape as well as in the velocity distribution because the inner region of the torus with the clouds of the high velocities can be seen in this orientation.

	It was discussed in Imanishi et al. (2018), in the framework of Keplerian motions in the torus of NGC1068 the orbital velocity of the clouds corresponds to the SMBH mass is $9 \times 10^5 M_\odot$. This SMBH mass value is much lower than the previous estimations. In considered dynamical model, the observational low velocity of the clouds  (20 km/s at the distance 3 pc) can be explained by  the SMBH mass $5 \times 10^6 M_\odot$. Such low velocity for more high value of the SMBH mass is related to the fact that clouds move in eccentric orbits due to self-gravity of the torus (Bannikova \& Sergeev 2017). It could mean that ALMA observations demonstrate the outer region of the torus where the clouds pass the apocenter of the orbits.

\section{Conclusions}
\label{Conclusions} 

The proposed dynamical model of a dusty torus takes into account the gravitational interaction between clouds which are moving in the field of the central mass. Such a model can explain the thickness of the torus which naturally appears due to the spread of the clouds in inclinations and in eccentricities. The distribution of the clouds in the torus cross-section obeys the Gaussian law, in agreement with the mid-IR spectra of AGNs. The velocity distribution in the torus of NGC1068 obtained by ALMA can be explained by orientation effects and clumpy structure. The best match of the model velocity map with the observed one occurs for an inclination of the torus of $35^\circ$ from edge-on position. The final orientation of the symmetry axis of the torus corresponds to the observed orientation of the wind cones. The SMBH mass in NGC1068 is about $5 \times 10^6 M_\odot$. This result was obtained by accounting for the torus self-gravity.

\section*{ACKNOWLEDGEMENTS}

\noindent We thank Massimo Capaccioli and Oleg Ulyanov for the fruitful discussions which helped us to improve this work.
This paper makes use of the following ALMA data: ADS/JAO.ALMA$\sharp$2016.1.00052.S. ALMA is a partnership of ESO (representing its member states), NSF (USA) and NINS (Japan), together with NRC (Canada), MOST and ASIAA (Taiwan), and KASI (Republic of Korea), in cooperation with the Republic of Chile. The Joint ALMA Observatory is operated by ESO, AUI/NRAO and NAOJ.
\subsection*{\rm \bf \normalsize References}

\setlength\parindent{-24pt}

\par
  
Antonucci R.R.J. (1984) ApJ 278, 499

Antonucci R.R.J. \& Miller J.S. (1985) ApJ 297, 621

Bannikova  E.Yu., Vakulik V.G., Sergeev A.V. (2012) MNRAS 424, 820 

Bannikova E.Yu. (2015) Radio Physics and Radio Astronomy  20, 191 (in Russian) 

Bannikova E.Yu. \& Sergeyev A.V. (2017) Frontiers in Astronomy and Space Sciences  4, id.60 

Das V., Crenshaw D.M., Kraemer S.B., Deo R.P. (2006) AJ 132, 620

Gallimore J. F., Baum S. A., O'Dea C. P., Brinks E., Pedlar A. (1996) ApJ 462, 740

Gallimore J. F., Baum S. A., O'Dea C. P. (2004) ApJ 613, 794

Garcia-Burillo S., Combes F., Ramos Almeida C., et al. (2016) ApJL 823, L12

Garcia-Gonzalez J., Alonso-Herrero A., Honig  S. F., Hernan-Caballero A., Ramos Almeida C., Levenson N. A., et al. (2017) MNRAS 470, 2578

Gravity Collaboration, Sturm E., Dexter J., et al. (2018)  Nature 563, 657
 
Greenhill L. J., Gwinn C. R., Antonucci R. et al. (1996), ApJL 472, L21

Imanishi M., Nakanishi K., Izumi T., Wada K. (2018) ApJL 853, L25

Hur\'e J.-M. (2002) A\&A, 395, L21

Jaffe W., Meisenheimer K., Rottgering H. J. A. et al. (2004) Nature 429, 47

Krolik J. H. \& Begelman M. C. (1988) ApJ 329, 702

Lodato G. \& Bertin G. (2003) A\&A 398, 517

Nenkova M., Sirocky M. M., Ivezic Z., Elitzur M., (2008a) ApJ 685, 147

Nenkova M., Sirocky M. M., Ivezic Z., Elitzur M., (2008b) ApJ 685, 160

Osterbrock D. E. \& Shaw R. A. (1988) ApJ 327, 89

Schartmann, M., Burkert, A., Krause, M., Camenzind, M., Meisenheimer, K., Davies, R. I. (2010) MNRAS 403, 1801

Tristram, K. R. W., Burtscher, L., Jaffe, W., Meisenheimer, K., Honig, S. F., Kishimoto, M., et al. (2014) A\&A 563, A82

Tristram, K. R. W., Meisenheimer, K., Jaffe, W., Schartmann, M., Rix, H.-W., Leinert, C., et al. (2007) A\&A 474, 837
 
Urry C. M. \& Padovani P. (1995) Publ. Astron. Soc. Pac. 107, 803 
 
\end{document}